\documentclass[preprintnumbers,amsmath,amssymb]{revtex4}

\usepackage{graphicx}
\usepackage{bm}
\usepackage{latexsym}
\usepackage{amssymb}
\usepackage{amsmath}
\newtheorem{theorem}{Theorem}[section]

\newtheorem{lemma}{Lemma}[section]
\newtheorem{assumption}{Assumption}[section]

\begin{document}

\title{Existence of breathers in nonlinear Klein-Gordon lattices}

\author{Dirk Hennig}
\affiliation{Department of Mathematics, University of Thessaly, Lamia GR35100, Greece}

\date{\today}

\begin{abstract}
\noindent 
We prove the existence of  time-periodic and spatially localised solutions (breathers)
in general  nonlinear Klein-Gordon  infinite  lattices of weakly coupled oscillators by using Schauder's fixed point theorem  establishing that there are  time-reversal initial conditions leading to breather solutions. 

\end{abstract}

\maketitle

\section{Introduction}\label{section:intro}

\noindent Discrete breather solutions in nonlinear lattices
have  attracted significant interest recently, not least due to the important role they play in many 
physical realms where features 
of localisation in systems of coupled oscillators are involved 
(for a review see \cite{Flach1} and references
therein),\cite{Lars}-\cite{Sakovich}. 
Proofs of existence and nonexistence
of breathers, as spatially localised and time-periodically varying solutions,   were 
provided in \cite{MacKay}-\cite{Dirk}. The exponential stability of breathers 
was proven in \cite{Bambusi}. Analytical and numerical methods have been
developed to continue breather solutions in conservative and dissipative systems  
starting from the anti-integrable limit \cite{Sepulchere}-\cite{Morgante}. 

Here we present a consice existence proof of breather solutions 
in general  nonlinear Klein-Gordon infinite lattices of weakly coupled oscillators using methods alternative to the approaches in \cite{MacKay}-\cite{Dirk}.  That is, we demonstrate that there exist initial conditions, complying with the time reversibility of the system,  such that the ensuing evolution is characterised by spatially localised and time-periodic solutions. In order to prove the  existence of such initial conditions, we formulate the problem  in terms of  an operator equation on a function space  
which is solved by virtue of {\bf Schauder's Fixed Point Theorem:} 
Let $S$ be a closed convex subset of the Banach space $X$. Suppose $g: S \rightarrow S$ is compact. Then $g$ has at least one  fixed point in $S$ (see in \cite{Schauder}-\cite{Zeidler}).

\noindent {\bf Definition:}  Let $X$ and $Y$ be normed spaces. The map $g: M\subseteq X \rightarrow Y$ is called {\it compact} iff\\
(i) $g$ is continuous, and \\
(ii) $g$ transforms bounded sets into relatively compact sets.

\vspace*{0.5cm}

We study the dynamics of general  nonlinear Klein-Gordon (KG) infinite lattice systems
given by the following set of coupled equations
\begin{eqnarray}
\frac{d^2q_n}{dt^2}&=&-V^{\prime}(q_n)+\kappa (q_{n+1}-2q_n+q_{n-1}),\,\,n \in {\mathbb{Z}},
\label{eq:start}
\end{eqnarray}
and the prime $^{\prime}$ stands for the derivative with 
respect to $q_n$, the latter being 
the coordinate of the oscillator at site $n$ evolving in an
anharmonic on-site potential 
$V(q_n)$.
Each oscillator interacts with its neighbours to the left and right with strength that is regulated by the value of the non-negative parameter $\kappa$.

In what follows we differentiate between soft on-site potentials and hard on-site potentials. For the former (latter)
the oscillation frequency of an oscillator moving in the on-site potential $V(q)$
decreases (increases) with increasing oscillation amplitude.

We make the following assumptions:

\vspace*{0.5cm}

\begin{assumption}\label{assumption1}
%{\bf Assumption (A1):} 
The anharmonic on-site potentials $V(x)$ are analytic and have the following properties:
\begin{equation}
  V(0)=V^{\prime}(0)=0\,,\,\,\,V^{\prime \prime}(0)=\omega_0^2> 0.
  \nonumber
%\label{eq:assumptions}
\end{equation}

Soft on-site potentials $V(x)$ possess adjacent to the minimum at $x=0$ 
inflection points at 
$x_{i}^{-}$ and maxima at $  x_M^{-}$, respectively with $x_M^+>x_i^+>0$ and $x_M^-<x_i^-<0$, 
and one has 
\begin{eqnarray}
V^{\prime}(x_{M}^-<x<0)&<&0,\,\,\,V^{\prime}(x_{M}^{-})=0,\,\,\,  V^{\prime}(0<x<x_{M}^+)>0,%
\nonumber
\label{eq:U1prime},\\
V^{\prime \prime}(x_{i}^-<x<x_{i}^+)&>&0,\,\,\,V^{\prime \prime}(x_{i}^{-})=0,\,\,\,
V^{\prime \prime}(|x_{i}^{-}|<|x|<|x_{M}^{-}|)<0.
\nonumber
%\label{eq:U2prime}
\end{eqnarray}

(We remark that $V(x)$ can have more than one (local) minimum. 
An example is a periodic potential $V(x)=1-\cos(x)$. 
However, in the frame of the current study we are only interested in motion 
between the maxima, $ x_M^{-}$, adjacent to the minimum of $V(x)$ at $x=0$. Furthermore, the case when a soft on-site potential has only one inflection point can be easily implemented. An example is $V(x)=(\omega_0^2/2)x^2-a x^3$, $a>0$.)  

\vspace*{0.5cm}

Hard on-site potentials $V(x)$ are
characterised 
by 
\begin{eqnarray}
V^{\prime}(x<0)&<&0,\,\,\,V^{\prime}(x>0)>0,
\nonumber
%\label{eq:assumptionhard}
\\ 
V^{\prime \prime}(x)&>& 0,\,\,\,\forall x \in \mathbb{R}.\nonumber
\end{eqnarray}

For soft and hard on-site potentials $V^{\prime}(x)$ can be expressed as
\begin{equation}
 V^{\prime}(x)=\omega_0^2x+W^{\prime}(x),\nonumber
\end{equation}
so that
\begin{equation}
V(x)=\frac{1}{2}\omega_0^2x^2+W(x),\nonumber
%\label{eq:Af}
\end{equation}
 with the anharmonic part $W$. 

 Further, we assume that some positive constants $\alpha$,$\beta$,$K_\alpha$ and $K_\beta$  the condition
\begin{equation}
|W^{\prime}(x)|\leq K_\alpha |x|^\alpha,\qquad |W^{\prime}(x)-W^{\prime}(y)|\leq K_\beta(|x|^\beta+|y|^\beta)|x-y|
	\label{eq:assumptions1}
\end{equation}
for all $x\in\mathbb{R}$  for hard on-site potentials, and for $x_M^-\le x\le x_M^+$ for soft on-site potentials, 
is fulfilled.
\end{assumption}

\vspace*{1.0cm}

The system of equations (\ref{eq:start}) has an energy integral 
\begin{equation}
 E=\sum_{n \in {\mathbb{Z}}}\left[\frac{1}{2}\dot{q}_n^2+V(q_n) \right]+
 \frac{\kappa}{2} \sum_{n \in {\mathbb{Z}}}(q_{n+1}-q_n)^2,\nonumber
\end{equation}
relating it  to a Hamiltonian structure,
associated with the Hamiltonian,
\begin{equation}
 H=\sum_{n \in {\mathbb{Z}}}\left[\frac{1}{2}{p}_n^2+V(q_n) \right]+
 \frac{\kappa}{2} \sum_{n \in {\mathbb{Z}}}(q_{n+1}-q_n)^2,\label{eq:Hamiltonian}
\end{equation}
where $p_n$ and $q_n$ are canonically conjugate momentum and coordinate variables.
Denoting $p=(p_1,p_2,...)$ and $q=(q_1,q_2,...)$, we note that 
the Hamiltonian system is  
time-reversible with respect to the involution $p\mapsto -p$.

\vspace*{0.5cm}

For systems  (\ref{eq:start}) with a hard on-site potential for any finite initial data the 
solutions are always bounded, that is
\begin{equation}
 ||q(t)||_{l^\infty}\le q_E<\infty,\,\,\,\forall t>0,\label{eq:qM}
\end{equation}
where $q_E$ depends on the energy level $E>0$ in (\ref{eq:level}), 
% \begin{equation}
%  \sup_{\begin{array}{c}
%                                         n\in {\mathbb{Z}}, t\in \mathbb{R}\\
%                                     
%                                        \end{array}}|q_n(t)|<\infty.\label{eq:qM}
%                                        \end{equation} 
because 
for all values of the 
total energy $0<E< \infty$ there exists 
a closed maximum equipotential surface 
\begin{equation}
\Sigma:\,\,\, \sum_{n \in {\mathbb{Z}}} V(q_n)+\frac{\kappa}{2} \sum_{n \in {\mathbb{Z}}}(q_{n+1}-q_n)^2=E,\label{eq:level}
\end{equation}
with ${p}_{n}\,=0$ for all $n \in {\mathbb{Z}}$. As trajectories cannot penetrate  the maximum 
equipotential surface all coordinates $q(t)$ are bounded to 
perform motions about the only equilibrium position in configuration space at $q=0$. 
In contrast, 
for systems with a
soft on-site potential even for finite initial data (respectively sufficiently high values of $E$) 
unbounded solutions may ensue. An example for such a potential is $V(x)=(\omega_0^2/2)x^2-a x^4$, $a>0$.
Being interested only in bounded motion, we make the following assumption:

\vspace*{0.5cm}

\begin{assumption}\label{assumption4} For systems with a soft on-site potential assume that $E$ is chosen such 
that 
a closed maximum equipotential surface $\Sigma$ exists bounding all motions about 
the equilibrium position in configuration space situated at $q=0$ such that 
\begin{equation}
 ||q(t)||_{l^\infty}\le q_E\le q_M<\infty,\,\,\,\forall t>0,\label{eq:qMs}
\end{equation}
\end{assumption} 
The level of $E$ is
unrestricted for hard on-site potentials and is restricted to $(0, E_0)$ for some $E_0 < \infty$ for soft on-site potentials. 

We study the existence  of solutions to the system (\ref{eq:start}) that are 
 time-periodic and spatially localised satisfying
 \begin{equation}
  q_n(t+T)=q_n(t),\,\,\, \lim_{n\rightarrow - \infty} |q_n|=0,\nonumber
 \end{equation}
with period $T$. 

The solutions of the system obtained when linearising  equations (\ref{eq:start}) around the equilibrium $q_n=0$, are superpositions of plane wave solutions (phonons) 
\begin{equation}
 q_n(t)= \exp(i(kn-\omega_0 t)),\nonumber
\end{equation}
with frequencies 
\begin{equation}
 \omega_0^2(k) =\omega_0^2+4\kappa\sin^2\left(\frac{k}{2}\right),\,\,\,k\in[-\pi,\pi].\nonumber
\end{equation}
These (extended) states disperse. Therefore, the  frequency $\omega_b$ of a  localised time-periodic solution must satisfy the non-resonance condition $\omega_b \neq |\omega_0(k)|/m$ for any integer $m\ge 1$. This requires 
$\omega_b^2 >\omega_0^2+4\kappa$ ($\omega_b^2<\omega_0^2$) for hard (soft) on-site potentials as a necessary condition for the existence of localised time-periodic solutions of system (\ref{eq:start}). Therefore, for localised  solutions  the values of their   
frequency of oscillations $\omega_b$ have to lie above (below) the upper (lower)  edge  
of the continuous (phonon) spectrum  determined by $\sqrt{\omega_0^2+4\kappa}$ 
($\omega_0$), respectively, which is only achieved by amplitude-depending  tuning of the frequency due to the presence of 
the nonlinear term $W^{\prime}(x)\not \equiv 0$.

The system (\ref{eq:start}) with its associated Hamiltonian (\ref{eq:Hamiltonian}) 
belongs to the  class of Hamiltonian systems that are 
even in their momentum  variables.
The forthcoming results regarding
the existence of localised periodic solutions to
(\ref{eq:start}) are supported by the following statement.

\vspace*{0.5cm}
\begin{lemma}
\label{lemma1}
{\it Consider the system (\ref{eq:start}) on the infinite lattice $n \in {\mathbb{Z}}$. Let assumption \ref{assumption1} hold. 
In addition, for systems with soft  on-site potential let assumptions  \ref{assumption4} hold.
Further, let $(p_n(t),q_n(t))_{n\in {\mathbb{Z}}}$, 
 be the solution to the time-reversible system
 \begin{equation}
 \dot{p}_n=-\frac{\partial H}{\partial q_n},\,\,\,\dot{q}_n=\frac{\partial H}{\partial p_n},\,\,\,n\in {\mathbb{Z}}, \label{eq:system},
\end{equation}
with $H$ given in (\ref{eq:Hamiltonian}) and 
 with initial data
 \begin{equation}
 \left( p_n(0) \right)_{n \in \mathbb{Z}}=  0,\,\,\, \left( q_{n}(0)\right)_{n \in \mathbb{Z}} \neq 0,
 %\,\,\,n\in {\mathbb{Z}},
 \label{eq:icp} 
 \end{equation}
 and $(q_{n0})_{n \in \mathbb{Z}}\in l^1$.
% for which $q_{k0}\ne q_{l0}$ for at least one pair of indices $l\ne k$.
If for some $\tilde{t} > 0$ one has 
 \begin{equation}
  p_{n}(\tilde{t})  =p_{n}(0)=0,
  %\,\,\,|p_n(t)|\neq 0,\,\,\,0<t<T,
  \,\,\, n\in {\mathbb{Z}},\label{eq:periodlemma}
 \end{equation}
then 
the data $(p_{n}(0),q_{n}(0))_{n\in {\mathbb{Z}}}$ belong to a periodic orbit with period $2\tilde{t}$.
}
\end{lemma}
 
 \vspace*{0.5cm}

\noindent{\bf Proof:} Note that as a consequence of the assumptions on the 
initial data
(\ref{eq:icp}) in conjunction with the conservation of energy, $E\neq 0$, the attributed solution $(p_{n}(t),q_{n}(t))_{n\in {\mathbb{Z}}}$ 
to system (\ref{eq:system}) 
is non-trivial. Suppose there is a $\tilde{t}>0$ such that for initial data 
(\ref{eq:icp})  the  solution 
 satisfies the conditions in 
(\ref{eq:periodlemma}). Then the  
time-reversibility symmetry $p\rightarrow -p$ 
% in conjunction with the reversion symmetry $q\rightarrow -q$ of the 
% Hamiltonian it holds that $F(-X)=-F(X)$ leaving the system (\ref{eq:system}) 
% invariant under 
% $X \leftrightarrow -X$ being tantamount to invariance  
% with respect to a simultaneous change of the sign 
% of $p$ and $q$. This 
implies that the 
solution for the initial data (\ref{eq:icp}) possesses the following symmetry features
\begin{equation}
{p}_n(\tilde{t}+t)=-p_n(\tilde{t}-t),\,\,\,q_n(\tilde{t}+t)=q_n(\tilde{t}-t),\,\,\,0\le t\le \tilde{t},\,\,\,n\in {\mathbb{Z}}.\nonumber
\end{equation}
Therefore, 
\begin{equation}
q_n(0)=q_n(2\tilde{t}),\,\,\,p_n(0)=-p_n(2\tilde{t})=0,\,\,\,n\in {\mathbb{Z}},\nonumber
\end{equation}
verifying that the solution is  periodic with period $2\tilde{t}$ and the proof is complete.

\hspace{16.5cm} $\square$

\vspace*{1.0cm}

\section{Localised solutions for an infinite lattice of coupled nonlinear oscillators}\label{section:two}

In the following we prove the existence of spatially localised and time-periodic solutions 
for the system\,(\ref{eq:start}).  
% The localised solutions, like  normal modes, are understood as a 
% {\it vibration in unison} of the system, i.e. 
% all oscillators of the system perform synchronous oscillations 
% so that the coordinates and the momenta of all oscillators  attain their respective extreme values simultaneously. 

Using Duhamel's principle, the solution of system (\ref{eq:start}) with initial data
\begin{equation}
 \left( q_n(0)\right)_{n \in \mathbb{Z}}=\left(A_n\right)_{n \in \mathbb{Z}}
 %\not \equiv 0
 ,\,\,\,\left(\dot{q}_n(0)\right)_{n \in \mathbb{Z}} \equiv 0,
 %\,\,\,n\in {\mathbb{Z}},
 \label{eq:ics}
\end{equation}
can be expressed as a system of integral equations 
\begin{equation}
q_n(t)=A_{n}\,\cos(\omega_0 t)+\frac{1}{\omega_0}\,\int_0^t\,\sin[\omega_0
(t-\tau)]f(q_{n-1}(\tau),q_n(\tau),q_{n+1}(\tau))d\tau,\,\,\,n\in {\mathbb{Z}}\label{eq:ie1}
\end{equation}
where  $f(q_{n-1},q_n(\tau),q_{n+1})=-W^\prime(q_n)+\kappa(q_{n+1}+q_{n-1}-2q_n)$ and we assume
\begin{equation}
 A=\left(A_n\right)_{n\in {\mathbb{Z}}} \in l^1.\nonumber
 %\qquad \parallel A\parallel_{l^1}\le R.
\end{equation}
Note that by continuous embedding $A\in l^1 \subset l^2$ and $\parallel A \parallel_{l^2}\le \parallel A \parallel_{l^1}$.
To show the existence of a unique local solution of (\ref{eq:ie1}) we consider $q=(q_n)_{n\in {\mathbb{Z}}} \in C^1([0,s];l^2)$, where $s$ is a fixed positive number. Consider the  Banach space 
\begin{equation}
 {\cal{B}}=\left\{\,q\in C^1([0,s];l^2)\,\right\}\nonumber
\end{equation}
with norm 
\begin{equation}
 \parallel q \parallel_{\cal{B}}=\max_{t\in [0,s]}\left\{\parallel q(t)\parallel_{l^2}, \parallel \dot{q}(t)\parallel_{l^2}\,\right\}\nonumber
\end{equation}

We define a subset $S$ of ${\cal{B}}$ as
\begin{equation}
 S=\left\{\,q\in {\cal{B}}\,\,:\,\,\parallel q\parallel_{{\cal{B}}}\le 
R\,\right\}.\nonumber
%\label{eq:S}
\end{equation}
Clearly, $S$ is a closed,  bounded and convex set.

\vspace*{0.5cm}

Related to (\ref{eq:ie1}) we define an operator on $S$ as follows:
\begin{equation}
U(q)=A\,\cos(\omega_0 t)+\frac{1}{\omega_0}\,\int_0^t\,\sin[\omega_0
(t-\tau)]f(q(\tau))d\tau.\nonumber
\end{equation}

Next we show that the IVP (\ref{eq:start}),(\ref{eq:ics}) admits a unique global solution.

\begin{lemma}
\label{lemma2}
{\it For every $A\in l^1$ there exists a unique global solution $q(t)$ of the system (\ref{eq:start}) in $C^2([0,\infty),l^2)$ 
%for every $t \in {\mathbb{R}}$ 
such that $q(0)=A$ and $\dot{q}(0)=0$.} 
\end{lemma}

{\bf Proof:}
Local well-posedness of the initial value problem (IVP) (\ref{eq:start}),(\ref{eq:ics}) and differentiability of the local solution $q$ with respect to $t$ is shown using the contraction mapping principle applied to the integral representation (\ref{eq:ie1}).
Using the Banach algebra property of $l^2$, i.e. $||xy||_{l^2}\le ||y||_{l^2}\,||y||_{l^2}$ for all $x,y \in l^2$, 
we derive for the following upper bound 
\begin{equation}
 ||U(q)||_{B}\le\max\left\{1,\omega_0\right\}\left( R_0+\frac{s}{\omega_0}\left(K_\alpha R^\alpha+4\kappa R \right)\right),\,\,\,R_0=||A||_{l^2},\,\,\,\forall q \in S.\nonumber
\end{equation}

We may choose $R_0<R/2$ and $s\le \max\left\{1,\omega_0\right\}\cdot R\omega_0/(2(K_\alpha R^\alpha+4\kappa R))$, so that $U: S\mapsto S$.
Now for every $x,y\in S$ one has 
\begin{eqnarray}
 (U(x))_n(t)-(U(y))_n(t)&=&\frac{1}{\omega_0}\,\int_0^{t}\,\sin[\omega_0
(t-\tau)]\left(f(x_{n-1}(\tau),x_n(\tau),x_{n+1}(\tau))-f(y_{n-1}(\tau),y_n(\tau),y_{n+1}(\tau))\right)d\tau\nonumber\\
&=&\frac{1}{\omega_0}\,\int_0^{t}\,\sin[\omega_0
(t-\tau)]\left( (\Delta x)_n(\tau) -(\Delta y)_n(\tau)-(W^\prime(x_n(\tau)-W^\prime(y_n(\tau))\right)d \tau,\nonumber
\end{eqnarray}
and we used the notation $(\Delta x)_n=x_{n+1}+x_{n-1}- 2x_n$.
We derive the estimate
\begin{equation}
 ||U(x)-U(y)||_{B}\le \frac{s}{\omega_0}\left(4\kappa R+2K_\beta R^\beta\right)||x-y||_B.\nonumber
\end{equation}
Hence, by taking 
\begin{equation}
 s<\min\left\{ \frac{\omega_0}{4\kappa R+2K_\beta R^\beta}, \frac{R \omega_0}{2(K_\alpha R^\alpha+4\kappa R)}\right\},\nonumber
\end{equation}
one concludes that $U$ constitutes a contraction mapping on $S$ so that by the Banach Fixed Point Theorem there exists a unique fixed point of $U$ in $S$ which is a unique local solution of (\ref{eq:ie1}).

A maximal solution can be constructed by repeated application of the procedure above with initial data $A(s-s_0)$ for some $0<s_0<s$ where we exploit the uniqueness of the solution to continue the latter.

From (\ref{eq:start}) we get
\begin{equation}
 \sup_{t\in [0,s]}||\ddot{q}(t)||_{l^2}\le (K_\alpha R^{\alpha-1}+4\kappa)R,\nonumber
\end{equation}
assuring that the solution belongs to $C^2([0,s],l^2)$.

Global continuation of the solution is guaranteed by the energy method leading to the constraints  (\ref{eq:qM}) and (\ref{eq:qMs}) for hard and soft on-site potentials, respectively, from which follows that there is no blow-up of the solutions in finite time.

\hspace{16.5cm} $\square$

 \vspace{0.5cm}

 The following lemma on the smooth dependence of the solutions on the initial conditions will be made use of for the existence proof of breather solutions.
 \begin{lemma}\label{lemmaGronwall}
   Let  $\phi(t,0,A)$ be the solutions to the IVP determined by (\ref{eq:start}) and (\ref{eq:ics}) with
  $\sup_{t\in[0,\infty)}||\phi(t,0,A)||_{l^1}\le \overline{M}<\infty$.
  Then for all $A \in l^1$ it holds
  \begin{equation}
   ||\phi(t,0,A)||_{l^1}\le ||A||_{l^1}\,
   \exp\left[\frac{K_\alpha \overline{M}^{\alpha-1}+4\kappa}{\omega_0} t \right],\,\,\,\forall t\in [0,\infty).\label{eq:Gronwall}
  \end{equation}
 \end{lemma}
 
 {\bf Proof:} Using (\ref{eq:assumptions1})  and (\ref{eq:ie1}) we derive
 \begin{eqnarray}
  ||\phi(t,0,A)||_{l^1}&=&\sum_{n\in \mathbb{Z}}|A_{n}\,\cos(\omega_0 t)+\frac{1}{\omega_0}\,\int_0^t\,\sin[\omega_0
 (t-\tau)]f(\phi_{n-1}(\tau,0,A),\phi_n(\tau,0,A),\phi_{n+1}(\tau,0,A))d\tau |\nonumber\\
 &\le& ||A||_{l^1}+\frac{1}{\omega_0}\,\int_0^t\,\sum_{n\in \mathbb{Z}}|-W^\prime(\phi_n(\tau,0,A))+\kappa[\phi_{n+1}(\tau,0,A)- 2\phi_n(\tau,0,A)+\phi_{n-1}(\tau,0,A)]|d\tau\nonumber\\
 &\le& ||A||_{l^1}+\frac{K_\alpha \overline{M}^{\alpha-1}+4\kappa}{\omega_0}\,\int_0^t\,||\phi(\tau,0,A)||_{l^1} d\tau,\,\,\,\forall t\in [0,\infty).\nonumber
 \end{eqnarray}
 Facilitating Gronwall's inequality we get (\ref{eq:Gronwall}) concluding the proof.
 
 \hspace{16.5cm} $\square$
 
 \vspace{0.5cm}
 
Furthermore, we need the following result for compact operators on infinite dimensional Banach spaces.

\begin{lemma}
 \label{lemma:surjectice}
 Let $X$ be an inifinite dimensional Banach space and $g:\,X\mapsto X$ compact. Then $g$ is not surjective, i.e. there is $y\in X$ such that $g(x)=y$ has no solution.
\end{lemma}

{\bf Proof:} 
Denote by $B(0,r)=\{\,x\in X\,:\,||x||_X<r\,\}$ the open ball in $X$ of center $0$ and radius $r$.  For a contradiction, suppose that $g$ is surjective. By the open mapping theorem it follows that $g$ is an open operator. Particularly, $g(B(0,1))\subseteq X$ is an open set. That is, there exists $\epsilon>0$ such that 
\begin{equation}
{B}(0,\epsilon)\subseteq g(B(0,1)). \label{eq:closed}
\end{equation}
$g$ being compact, implies $g(B(0,1))$ is relatively compact. Then due to (\ref{eq:closed}), it follows that ${B}(0,\epsilon)$ is relatively compact. Furthermore, there exists  $\tilde{\epsilon}<\epsilon$, such that the closed ball $\overline{B}(0,\tilde{\epsilon})\subset g(B(0,1))$ and  $\overline{B}(0,\tilde{\epsilon})$ is compact.
However, if $\overline{B}(0,\tilde{\epsilon})$ is compact, then by the Riesz theorem $X$ must be finite dimensional, contradicting the hypothesis that $X$ is infinite dimensional. Therefore, $g$ is not surjective.
\\  $\Box$
 
\vspace*{0.5cm}

Finally, for the existence of breathers in system (\ref{eq:start}) we have the following statement:
\begin{theorem}
\label{maintheorem}
{%Let assumption \ref{assumption:A2} hold. 
Let $\mu=\max_{x\in I}|\sin(x)|/x=|\sin(u)|/u$ with $I=[0,\pi]$ ($I=[\pi,2\pi]$) for hard (soft) on-site potentials, where $u$ solves  $u=\tan(u)$ in the respective interval.

Consider  the solutions $\phi(t,0,A)$ to the IVP determined by (\ref{eq:start}) and (\ref{eq:ics}).
  Take initial data $A \in S_A \subset l^1$, where $S_A$ is the  closed ball  centered at $0$ of radius $R_A$ of $l^1$,
\begin{equation}
 S_A=\left\{\,A\in l^1\,\,:\,\,\parallel A\parallel_{l^1}\le 
R_A<\left( \frac{\omega_0^2 \mu}{K_\alpha}\right)^{1/(\alpha-1)}\,\right\}.\label{eq:constraint}
%\label{eq:SA}
\end{equation}
% Let 
% \begin{equation}
%  R_A<\left( \frac{\omega_0^2 \mu}{K_\alpha}\right)^{1/(\alpha-1)},\label{eq:constraint}
% \end{equation}
%  where 
%  $\mu=\sin(x)/x$  solves  $x=\tan(x)$ 
Then for $\kappa$ small enough there exists a breather solution.}
\end{theorem}

{\bf Proof:} We have 
\begin{equation}
\dot{\phi}_n(t,0,A)=-\omega_0 A_{n}\,\sin(\omega_0 t)+\int_0^t\,\cos[\omega_0
(t-\tau)]f(\phi_{n-1}(\tau,0,A),\phi_n(\tau,0,A),\phi_{n+1}(\tau,0,A))d\tau,\,\,\, n\in {\mathbb{Z}}.\nonumber
%\label{eq:momentum} 
\end{equation}
  
The relation $\dot{\phi}_n(\tilde{t})=0$ for all $n\in \mathbb{Z}$, for some $\tilde{t}$ with 
  $0<\tilde{t}<\pi/\sqrt{\omega_0^2+4\kappa}$ for hard on-site potentials  and $\pi/\omega_0<\tilde{t} <2\pi/\omega_0$, 
 for soft on-site potentials, is equivalent to
\begin{equation}
  A_{n}\,=-\frac{1}{\omega_0\sin(\omega_0 \tilde{t})}\,\int_0^{\tilde{t}}\,\cos[\omega_0
 (\tilde{t}-\tau)]f(\phi_{n-1}(\tau,0,A),\phi_n(\tau,0,A),\phi_{n+1}(\tau,0,A))d\tau,\,\,\, n\in {\mathbb{Z}}.\label{eq:equiv}
 \end{equation}  
The right side of system (\ref{eq:equiv}) constitutes a mapping $g:l^1 \rightarrow l^1$,
with 
\begin{equation}
 (g(A))_n=g_n(A)=-\frac{1}{\omega_0\sin(\omega_0 \tilde{t})}\,\int_0^{\tilde{t}}\,\cos[\omega_0
 (\tilde{t}-\tau)]
 f(\phi_{n-1}(\tau,0,A),\phi_{n}(\tau,0,A),\phi_{n+1}(\tau,0,A))
 d\tau,\,\, n\in {\mathbb{Z}}.\nonumber
\end{equation}

We show that $g$ is sequentially continuous on $l^1$. Let $(A_k)_{k \in {\mathbb{N}}}\subseteq l^1$ be such that $A_k \rightarrow A$ as $k \rightarrow \infty$ in $l^1$.
By the continuous dependence of the solutions on the initial conditions (by virtue of Lemma \ref{lemmaGronwall}), it holds that $\phi_{i}(\tau,0,A^k)\rightarrow \phi_{i}(\tau,0,A)$ as $k \rightarrow \infty$ for all $i \in \mathbb{Z}$ and $\tau \in [0,\tilde{t}]$. Furthermore, as $W^\prime(\phi_{i}(\tau,0,A^k)) \rightarrow W^\prime(\phi_{i}(\tau,0,A))$ and $(\Delta \phi(\tau,0,A^k))_i \rightarrow (\Delta\phi(\tau,0,A))_i$ as $k\rightarrow \infty$ for all $i \in \mathbb{Z}$ and $\tau \in [0,\tilde{t}]$, one has  
$f(\phi_{i-1}(\tau,0,A^k),\phi_{i}(\tau,0,A^k),\phi_{i+1}(\tau,0,A^k))\rightarrow f(\phi_{i-1}(\tau,0,A),\phi_{i}(\tau,0,A),\phi_{i+1}(\tau,0,A))$ as $k \rightarrow \infty$ for all $i \in \mathbb{Z}$ and $\tau \in [0,\tilde{t}]$. Hence,
\begin{equation}
 g_i(A^k) \rightarrow g_i(A)\,\,\,{\rm as}\,\,k\rightarrow \infty,\,\,\, \forall i \in \mathbb{Z},\nonumber
\end{equation}
 i.e. $g$ is sequentially continuous on $l^1$, and thus, as sequential continuity is equivalent to continuity in normed spaces, $g$ is continuous.

Obviously, $S_A$ is a closed,  bounded and convex set of $l^1$ and we consider the map $g: S_A \mapsto S_A$.
A solution of the 
system of equations (\ref{eq:equiv})  
is then a  fixed point of the operator equation
\begin{equation}
 A=g(A).\label{eq:op}
\end{equation}
In order to apply Schauder's Fixed Point Theorem to show the existence of at least one 
solution to (\ref{eq:op}), 
one needs to verify that the operator $g$ maps  $S_A$ into itself and is  
compact  on $S_A$. 

We introduce $M=\max_
 {{\begin{array}{l} t\in [0,\tilde{t}\,]\\ A \in B_{R_A} \end{array}}} ||\phi(t,0,A)||_{l^1}$,
and show that for initial data $A\in S_A=B_{R_A} \subseteq l^1$, with $R_A$ constrained by (\ref{eq:constraint}), 
and coupling strength $\kappa$ small enough, the relation 
 \begin{equation}
 \frac{K_\alpha M^{\alpha}+4\kappa M}{\omega_0|\sin(\omega_0 \tilde{t})|}\tilde{t}
  < R_A\label{eq:hyp2}
 \end{equation}
is fulfilled. 
Note that for the solutions, $\phi(t,0,A)$, to the IVP determined by (\ref{eq:start}) and (\ref{eq:ics}) it holds 
for $\kappa=0$ that $\max_{t\in[0,\tilde{t}]}||\phi(t,0,A)||_{l^1}=||A||_{l^1}$, and thus, $M=\max_
{{\begin{array}{l} t\in [0,\tilde{t}\,]\\ A \in B_{R_A} \end{array}}} ||\phi(t,0,A)||_{l^1}=R_A$. Therefore, when  $\kappa=0$,  (\ref{eq:hyp2}) is satisfied if  
\begin{equation}
\frac{K_\alpha R_A^{\alpha}}{\omega_0|\sin(\omega_0 \tilde{t})|}\tilde{t}
  < R_A,\nonumber
  \end{equation}
  yielding (\ref{eq:constraint}), that is
\begin{equation}
  R_A <\left(\frac{\omega_0|\sin(\omega_0 \tilde{t})|}{K_\alpha \tilde{t}}\right)^{1/(\alpha-1)}\le \left( \frac{\omega_0^2 \mu}{K_\alpha}\right)^{1/(\alpha-1)} .\label{eq:constraint1}
\end{equation}
By the continuous dependence of the solutions on the parameter $\kappa$, there is a sufficiently small $\kappa_0$ such that for $0<\kappa <\kappa_0$ the relation (\ref{eq:hyp2}) holds for all $A\in B_{R_A}$.
 
Using (\ref{eq:hyp2}) 
% and the continuous embeddings  $l^p\subset l^q,\,\,\,|| x||_{l^q}\le || x||_{l^p},\,\,\,1 \le p\le q \le \infty,$ 
we estimate
 \begin{eqnarray}
||g(A)||_{l^1}&=&\sum_{n \in {\mathbb{Z}}} |(g(A))_n|=\sum_{n \in {\mathbb{Z}}} \left|\frac{1}{\omega_0\sin(\omega_0 \tilde{t})}\,\int_0^{\tilde{t}}\,\cos[\omega_0
 (\tilde{t}-\tau)]
 f(\phi_{n-1}(\tau,0,A),\phi_{n}(\tau,0,A),\phi_{n+1}(\tau,0,A))
 d\tau\right|\nonumber\\
%  &\le& \frac{1}{\omega_0|\sin(\omega_0 \tilde{t})|}\,\int_0^{\tilde{t}}\,\sum_{n \in {\mathbb{Z}}} \left| f(\phi_{n-1}(\tau,0,A),\phi_{n}(\tau,0,A),\phi_{n+1}(\tau,0,A))\right|d\tau\nonumber\\
 &\le& \frac{1}{{\omega_0|\sin(\omega_0 \tilde{t})|}}\,\int_0^t\,\sum_{n\in \mathbb{Z}}|-W^\prime(\phi_n(\tau,0,A))+\kappa[\phi_{n+1}(\tau,0,A)- 2\phi_n(\tau,0,A)+\phi_{n-1}(\tau,0,A)]|d\tau\nonumber\\
 &\le& \frac{1}{{\omega_0|\sin(\omega_0 \tilde{t})|}}\,
 (K_\alpha M^{\alpha-1} +4\kappa)\int_0^t\, \sum_{n\in \mathbb{Z}}|\phi_n(\tau,0,A)|d\tau \nonumber\\
 &=&\frac{1}{{\omega_0|\sin(\omega_0 \tilde{t})|}}\,
 \,
 (K_\alpha M^{\alpha-1} +4\kappa)\int_0^t\,||\phi(\tau,0,A)||_{l^1} ^ d\tau\nonumber\\
 &\le& \frac{K_\alpha M^{\alpha}+4\kappa M}{\omega_0|\sin(\omega_0 \tilde{t})|}\tilde{t}< R_A,\nonumber
 \end{eqnarray}
 showing that for $\kappa$ sufficiently small indeed $g(S_A)\subseteq S_A$.
 
 In order to prove that $g$ is compact we consider a sequence $(A_k)_{k \in \mathbb{N}} \subseteq S_A\subseteq l^1$.
 %$k \in \mathbb{Z}$. 
 Since for every $k \in \mathbb{N}$ one has $||g(A_k)||_{l^1}\le R_A$, the sequence $g(A_k)$ is bounded. Hence, $g(A_k)$  possesses a weakly convergent subsequence (not relabeled) that converges to a $B\in l^1$  ($g(A_k) \rightharpoonup B$ as $k \rightarrow \infty$). Since $l^1$ possesses the Schur property, i.e. weak and norm sequential convergence coincide in $l^1$, 
$g(A^k)$ possesses a subsequence that strongly (norm) converges in $l^1$. Hence, $g$ maps bounded subsets of $l^1$ into relatively compact subsets of $l^1$ and therefore, is compact.
By Schauder's Fixed Point Theorem the system (\ref{eq:op}) has at least one solution $A \in l^1$.

It remains to verify that there is at least one nontrivial fixed point solution. By contradiction: To this end assume that 
for every $A\in S_A\setminus \{0\}\subseteq l^1$ there exists a $B\in l^1$, $B \neq 0$ solving the inhomogeneous system
\begin{equation}
 A-g(A)=B\neq 0.\nonumber
\end{equation}
Suppose that the kernel of the operator $g-I$ is trivial. Then, 
for every $A \in S_A\setminus \left\{ 0\right\}$, there is  $B\in l^1\setminus \left\{ 0\right\}$ solving the inhomogeneous system
\begin{equation}
	\label{contr0a}
 g(A)-A=B\neq 0.
\end{equation}
This is equivalent to $g(A)=A+B$ for all $A \in S_A\setminus \left\{ 0\right\}$, 
Since $||g(A)||_{l^1}=||A+B||_{l^1}\le R_A$ and $g: S_A\subseteq l^1\mapsto S_A \subseteq l^1$, we have that $A+B$ in $S_A$, requiring that $g$ is surjective. But as $S_A$ is infinite dimensional and $g:\,S_A\subseteq l^1\mapsto S_A\subseteq l^1$ is compact, from Lemma \ref{lemma:surjectice} follows that $g$ is not surjective. That is,   there is $A+B$ for which $g(A)=A+B$ has no solution which contradicts (\ref{contr0a}). Conclusively, the kernel of $g-I$ is not trivial so that there is $A\neq 0$ solving $g(A)-A=0$. Hence, the fixed point equation (\ref{eq:op})  possesses at least one non-trivial solution.

%\ \ $\Box$

In conclusion, for  $\tilde{t}>0$,  
 satisfying (\ref{eq:hyp2}) and  initial conditions $A \in l^1$, the existence of solutions  
 for which 
 \begin{equation}
  \dot{\phi}_n(\tilde{t},0,A)=0,\,\,\,\forall n\in \mathbb{Z},\nonumber%  %\label{eq:zeros}
 \end{equation}
is verified. Periodicity  of $\phi(t,0,A)$ with $\phi(0,0,A)=\phi(2\tilde{t},0,A)$, is a consequence of the assertions of  Lemma \ref{lemma1}. 

% Due to the continuous embedding $A\in l^1 \subset l^2$ and $\parallel A \parallel_{l^2}\le \parallel A \parallel_{l^1}$, 
Since  $\phi(t,0,A)\in C^2([0,\infty);l^2)$
the localised solutions on the infinite lattice ${\mathbb{Z}}$ are represented 
by (infinite) square-summable sequences, viz. decay of the states for $|n|\rightarrow \infty$ takes place in the sense of the 
%$l^1$ respectively 
$l^2$ norm. In order to establish the existence of single-site breathers employing the above fixed point method, appropriately (e.g. exponentially) weighted function spaces can be used (see in \cite{JMP}).

\hspace{16.5cm} $\square$

\vspace*{0.5cm}
Final remarks: Notice that (\ref{eq:constraint1}) implies, when the frequency of the breather, $\omega_b=4\pi/\tilde{t}$, approaches the edge of the continuous (phonon) spectrum, that is  $\omega_b \rightarrow \sqrt{\omega_0^2+4\kappa}^{\,+}$ and small $\kappa$ for hard on-site potentials, and   $\omega_b \rightarrow \omega_0^{\,-}$ for soft on-site potentials, then  $R_A \rightarrow 0$, that is the amplitude of the breather goes to zero.
It is certainly of interest to extend the fixed point method to prove the existence of breathers also in higher dimensional lattices $Z^{d>1}$. 
 
\vspace*{0.5cm}

 \centerline{{\bf Acknowledgement}}
 
 I am very grateful to Nikos I. Karachalios for many stimulating discussions.

\end{document}